\documentclass[referee]{aa} 
\usepackage{graphicx}
\usepackage{txfonts}
\usepackage{natbib}
\bibpunct{(}{)}{;}{a}{}{,}
\begin{document}
\authorrunning{Janardhan, Tripathi \& Mason} 
\titlerunning{Source Region Evolution of Disappearance Events} 
\title{The solar wind disappearance event of 11 May 1999: source region evolution }

\author{P. Janardhan\inst{1}, D. Tripathi\inst{2}, and H.~E. Mason\inst{2}}

\offprints{jerry@prl.res.in}

\institute{${^{1}}$Physical Research Laboratory, Astronomy \& Astrophysics Division, 
              Navrangpura, Ahmedabad -- 380 009, India \\
	      \email{jerry@prl.res.in}\\
              ${^{2}}$Department of Applied Maths and Theoretical
              Physics, University of Cambridge, Wilberforce Road,
              Cambridge CB3 0WA, UK \\ 
	      \email{[D.Tripathi; H.E.Mason]@damtp.cam.ac.uk}} 
 
\date{Received ; accepted}

\abstract {A recent, detailed study of the well-known ``{\it {solar
wind disappearance event}}" of 11 May 1999 traced its origin to a
coronal hole (CH) lying adjacent to a large active region (AR), AR8525 
in Carrington rotation 1949. The AR was located at central
meridian on 05 May 1999 when the flows responsible for this event
began.  We examine the evolution of the AR-CH complex 
during 5$-$6 May 1999 to study the changes that apparently played a 
key role in causing this disappearance event.} {To study the evolution of 
the solar source region of the disappearance event of 11 May 1999.} 
{Using images from the Soft X-ray Telescope (SXT), the Extreme-ultraviolet 
Imaging Telescope (EIT) and the Michelson Doppler Imager (MDI) to examine 
the evolution of the CH and AR complex at the source region of the 
disappearance event.} {We find a dynamic evolution 
taking place in the CH-AR boundary at the source region of the 
disappearance event of 11 May 1999.  This evolution, which is found to 
reduce the area of the CH, is accompanied by the formation of new loops 
in EUV images that are spatially-and-temporally correlated with emerging flux 
regions as seen in MDI data.}{In the period leading up to the disappearance 
event of 11 May 1999, our observations, during quiet solar conditions and 
in the absence of CMEs, provide the first clear evidence for Sun-Earth 
connection originating from an evolving AR-CH region located at central meridian.  
With the exception of corotating interacting regions (CIR), these observations 
provide the first link between the Sun and space weather effects at 1 AU, 
arising from non-explosive solar events.}


\maketitle

\section{Introduction} 

During the period 11-12 May 1999 solar wind densities, as observed by
the Advanced Composition Explorer spacecraft (ACE; \citet{StF98}) 
located upstream of the Earth at the Lagrangian point L1, dropped to
unusually low values ($<$ 1 cm${^{-3}}$) for extended periods of time
($>$ 24 hours). This unusual and extended density depletion was also
accompanied by very low-velocity solar wind flows ($<$300 km
s${^{-1}}$) and caused a dramatic expansion of the Earth's
magnetosphere and bow shock \citep{LeC00}. It has been estimated that
the expanding bow shock moved outwards to a distance of $\sim$ 60
Earth radii, the lunar orbit, from its normal location of $\sim$10
Earth radii.  The extremely spectacular nature of this event has
caused it to be referred to as the ``{\it{day the solar wind
nearly died}}" \citep{Loc01}.

In a recent study of this event, \cite{JaF05} traced the solar 
wind outflows, observed at 1 AU, back to the Sun and showed that the 
flows responsible for the event began on 05 May 1999 from an active 
region-coronal hole (AR-CH) complex located at central meridian.  
They suggested that a continuously evolving CH boundary could cause a 
pinch-off, leading in turn to a separation of the CH outflow.  The 
expansion of this detached solar wind flow, by a factor of 6-7, could 
then give the desired low densities at 1 AU.  Given that the travel time 
between the Sun and the Earth, at the low velocities observed, was 
$\sim$5$-$6 days, it was argued that a pinch-off taking place $\sim$24$-$48 
hours after the start of the coronal hole outflow would cause typical 
particle densities of 20$-$25 particles cm${^{-3}}$ at $\sim$0.5 AU 
(the approximate distance the CH outflow would have moved outwards in 
48 hours) to be reduced to 0.1 particle cm${^{-3}}$ at 1 AU.  Thus, the 
expansion of a large, detached, low-velocity flow region from a small 
CH (as it propagated out to 1 AU) could give rise to an extremely large, 
low-density cloud that engulfed the Earth on 11 May 1999, as seen from 
interplanetary scintillation (IPS) observations \citep{BaJ03}.  

Isotopic ratios of O${^{7+}}$/O${^{6+}}$ are known to be good proxies for 
associating solar wind outflows to either AR or CH \citep{LNZ04}.  In a recent 
study it was shown that the solar source of this event could not be pinned 
down to either an AR or a CH \citep{JaF08}, as the O${^{7+}}$/O${^{6+}}$ 
ratios were sometimes indicative of a CH origin and at other times indicative 
of an AR origin.  Such a fluctuating O${^{7+}}$/O${^{6+}}$ signature was attributed 
by \cite{JaF08} to a dynamic and rapid evolution taking place at the AR-CH 
boundary region, which was at the solar source of the event.

Although many studies have attempted to understand the disappearance event 
of 11 May 1999 \citep{CrS00, FaS00, RiB00, UGF00, BaJ03, JaF05, JaF08}, none 
have examined the source region of this event.  We study the
%
\begin{figure}
\centering
\includegraphics[width=0.85\textwidth]{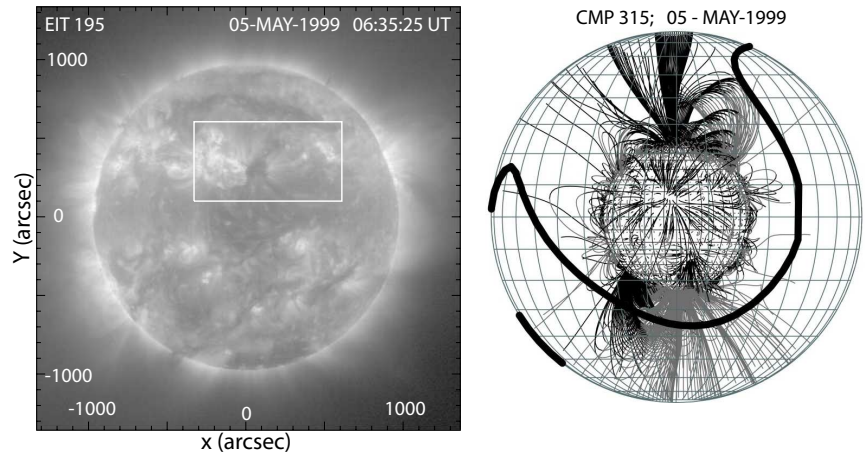}
\caption{Left: Full-disk EIT 195~{\AA} image on 05 May 1999.  The boxed 
region contains AR8525 and the small CH lying against its westernmost 
boundary.  Right: Three-dimensional structure of the coronal magnetic 
field on 05 May 1999, with the field lines projected on to a source 
surface at 2.5R${_{\odot}}$.  The thick wavy line is the magnetic 
neutral line.  See text for details.}
\label{fig1}
\end{figure}
source region of the 11 May 1999 disappearance event to understand 
the evolution and dynamics of this AR-CH complex and to try and pin-point its 
implications to solar-terrestrial relationships in the absence of explosive 
solar events.

\section{Observations}

The EIT \citep{DeA95} provides observation of the Sun at four different wavelengths
$\it{viz.~}$171~{\AA}~(\ion{Fe}{ix/x}; 1.0~MK), 195~{\AA}~(\ion{Fe}{xii}; 1.5~MK), 
284~{\AA}~(\ion{Fe}{xv}; 1.8~MK), and 304~{\AA}~(\ion{He}{ii}; 0.05~MK). The images 
recorded at 171~{\AA}, 195~{\AA}, and 284~{\AA} are mainly dominated by iron lines 
and probe systematically higher heights and temperature regions in the corona.  However, 
the images recorded at 195~{\AA} are contaminated with the \ion{Fe}{viii} line 
as well as the \ion{Fe}{xxiv} line, with the former being dominant in CH regions and 
later in flaring ARs \citep{DBM03, TrD06}. Therefore, these images can be used to 
study the evolution of the both ARs and CHs.  

In addition to EIT images, SXT images with the Al/Mg filter \citep{TsA91} were used 
to study the higher temperature (3$-$5 MK) responses and MDI line-of-sight magnetograms 
\citep{ScB95}, from the Solar and Heliospheric Observatory \citep{DFP95}, were used to 
study the evolution of photospheric magnetic fields around 05 May 1999, the 
approximate launch time of the disappearance event of 11$-$12 May 1999.  The images 
were processed using the standard IDL based solar-soft software tree.

\section{Data analysis and results}

Figure \ref{fig1} (left) shows a full disk EIT 195~{\AA} image on 05 May 1999.  The 
white box on the solar disk encloses the AR-CH complex, which is the region of 
interest in this study.  The small CH can be clearly seen butting up against the 
western most boundary of AR8525, which is located at central meridian. 
Figure~\ref{fig1} (right) shows the three-dimensional structure of the coronal 
magnetic field on 05 May 1999 as viewed from a Carrington longitude of 315${^{\circ}}$, 
the central meridian passage longitude for 05 May 1999.  The fields were computed using a 
potential field source surface (PFSS) model \citep{HKo99}.  The 
differently-shaded magnetic field lines distinguish the two polarities (black: positive 
and grey: negative) and are shown projected onto a source surface at 
2.5 R${_{\odot}}$, beyond which the potential field lines are assumed to be 
radial.  Only fields between 5$-$250 G on the 
photosphere are plotted.  The thick wavy line is the solar magnetic neutral 
line.  The black, outward pointed open fields lines at central meridian and 
slightly north of the equator are clearly visible and correspond to the 
location of AR8525 and the CH.  Based on the PFSS model it is clear that 
the target region shows open field lines emanating from the AR-CH complex. 

\begin{figure}
\centering
\includegraphics[width=0.85\textwidth]{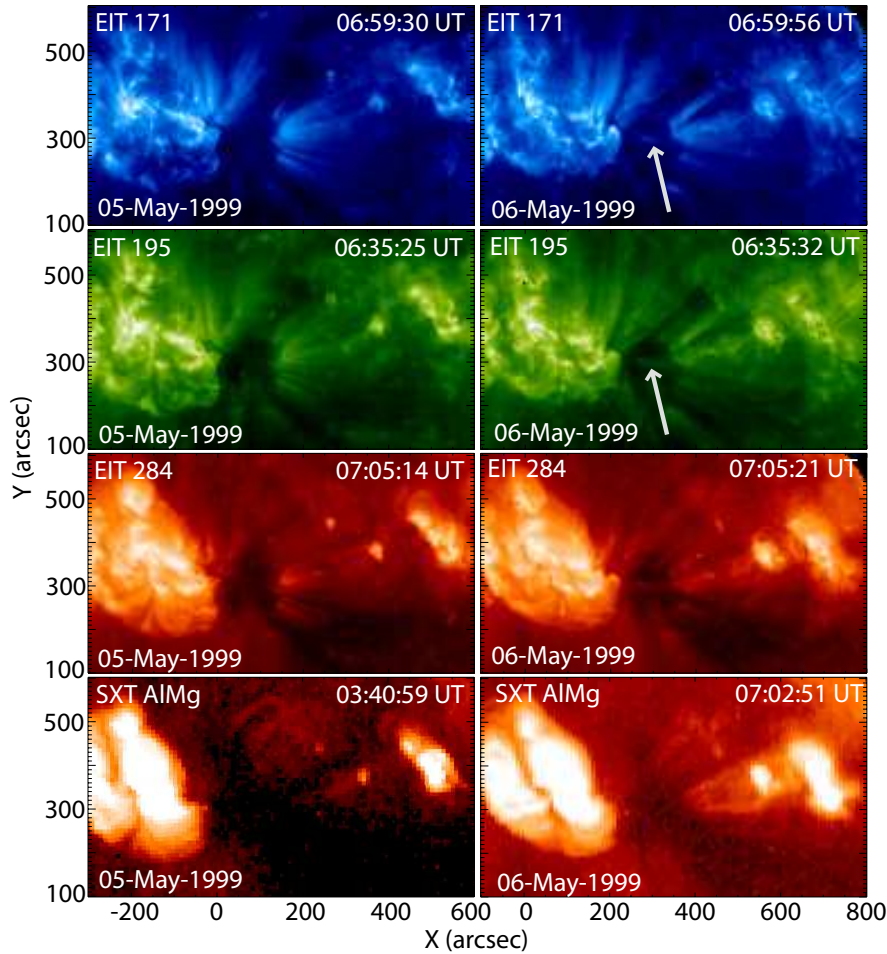}
\caption{The boxed region of the solar disk from Fig. \ref{fig1} 
on 05 May 1999 (left column) and 06 May 1999 (right column).  
From the top down are shown respectively, EIT 171~{\AA}; EIT 
195~{\AA} ; 284~{\AA}; and an SXT image. The white arrows in two of 
the right-hand panels point to new bright features in the CH.}
\label{fig2}
\end{figure}

Figure~\ref{fig2} shows images of the solar disk corresponding to the boxed 
region from the left-hand panel of Fig.~\ref{fig1}.  Each image is approximately 
centered on the AR-CH complex AR8525 on 05 May 1999 (left-hand panels) 
and 06 May 1999 (right-hand panels).  Starting from the top, the panels show 
respectively, EIT 171~{\AA}; EIT 195~{\AA}; EIT 284~{\AA}, and SXT 
images with the Al/Mg filter on 05 May 1999 (left) and 06 May 1999 (right). 
%
The small CH lying $\sim$300 arcsec north and immediately adjacent to 
AR8525, whose western-most boundary is located almost exactly at central 
meridian, can be easily identified in the images.  New bright features at the 
center can be seen to be producing a discernible change in the CH region on 06 May
1999, as compared with the previous day.  These changes, perceived as a 
constriction developing across the CH, are indicated by white arrows in two of 
the right-hand panels.  The two SXT images (lower most panels) 
also show a change in the emission on 06 May as compared to 05 May. It may 
be noted that images taken on 05 May and 06 May in Fig. \ref{fig2} are normalized to the 
same intensity scaling.  

\begin{figure}
\centering
\includegraphics[width=0.85\textwidth]{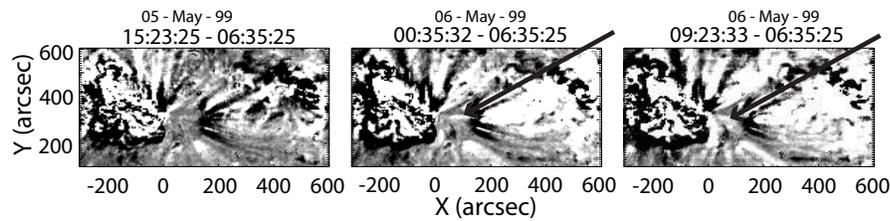}
\caption{Base difference images obtained at intervals of $\sim$9
hours.  In the middle and right-hand panels, the reference image 
at the left at 06:35:25 UT on 05 May 1999 has been subtracted from 
images $\sim$9 hours ahead of it.  The arrows in the middle and 
right-hand panels indicate the changes taking place in the CH.}
\label{fig3}
\end{figure}

To further substantiate the occurrence of this constriction or pinch-off taking 
place in the CH, we show EIT base difference images of the region in Fig. \ref{fig3}. 
The images were obtained by subtracting a reference EIT image on 05 May 1999 at 
06:35:25 UT from EIT images obtained at intervals of $\sim$9 hours ahead of the 
reference image.  Note that the images were differentially rotated to the time of the 
reference image.  The black regions in the difference images represent original features 
from the reference image, while the white regions show changes that have occurred from the 
time of the reference image.  The three panels clearly show the changes that can be 
seen to produce a constriction or narrowing of the CH in the $\sim$24 hour 
interval between the first panel on the left and the third panel on the right. 
The evolution of the CH, as observed by the EIT at 195~{\AA}, can be unambiguously 
seen in the base-difference movie Movie1.gif\footnote{Movies are available on line at 
http://www.edpsciences.org}, wherein the new bright features within the CH can 
be seen to be producing a progressive reduction in its area by causing a clear 
constriction or ``{\sl{pinch-off}}" across the CH.  
%

\begin{figure}
\centering
\includegraphics[width=0.85\textwidth]{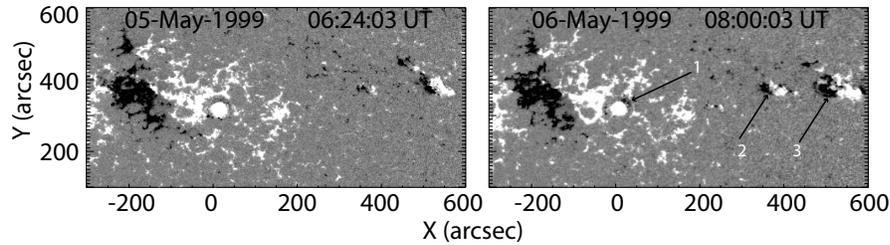}
\caption{MDI magnetograms of the boxed region of the solar disk 
from Fig. \ref{fig1}.  The panels differ in time by a 
little over 24 hours.  Arrow 1 shows a small region of newly 
emerging negative flux while arrows 2 and 3 show two 
evolving bipolar regions.}
\label{fig4}
\end{figure}

Figure \ref{fig4} shows MDI magnetograms, displayed between $\pm$300 G, 
of the boxed region of the solar disk from the left-hand panel of Fig.{\ref{fig1}}.  
The left-hand panel is at 06:24:03 UT on 05 May 1999 while the right-hand 
panel is on 06 May, a little over 24 hours ahead of this time.  
Note that the image in the right-hand panel has been rotated to 
the time of the panel on the left.  The black and white regions in each panel 
correspond to negative and positive polarities respectively.  The small, white, 
circular region of strong magnetic field lying slightly north and almost exactly 
at central meridian corresponds to the location of a small sunspot.  The negative 
polarities surrounding the strong sunspot field on 05 May are moving magnetic features 
that appear around spots during their decay phase \citep{HHa73}. On 06 May, a 
new negative polarity (shown by arrow numbered 1), whose corresponding positive 
polarity is not identifiable  unambiguously, is seen to appear to the northwest of 
the sunspot field.  Also seen are two bipolar regions to the far west (arrows numbered 2 
and 3), with the westernmost being clearly seen from 04 May and the one to 
its east beginning to emerge and evolve from 04 May.  The brightness observed 
in EIT and SXT images at these locations indicates the presence of hot closed loops.

The constriction taking place in the CH and seen as new bright features in its central 
region ( see Fig. \ref{fig2}) can take place by a process of interchange reconnection 
\citep{BVA07} wherein the open CH field lines reconnect with the closed field lines 
to the west.  This process will reduce the number of open field lines in the CH, thereby 
reducing the earth-directed solar wind outflows, produce the observed brightness at its 
center and shift the open field lines to a new location.  The 
new locations of these shifted open fields need not be ideally located to produce earth directed 
outflows and would therefore not contribute to the subsequent events at 1 AU.  The interchange 
reconnection process could also occur between the open CH fields and the closed field lines
anchored at one end at the new negative polarity seen to emerge on 06 May (marked by arrow 
numbered 1 in Fig. \ref{fig4}) or between the open CH fields and the closed loops at the 
two bipoles to the west.  A number of small new positive and negative polarities are also 
seen to appear around the CH location on 06 May.  It is therefore possible that the interchange 
reconnection process could initially start between the CH open fields and these new closed 
loops to initiate the constriction process and then lead up to interchange reconnection with 
closed loops at the bipole locations to the west, in a gradual and stepwise reconnection 
process \citep{AtH07, MaN07}.  For a detailed view of the evolution sequence in the 
AR-CH complex see the movie, Movie2.gif${^{1}}$.

\section{Discussion and Conclusions}

Using both spacecraft observations and tomographic IPS observations 
\cite{JaF05} located the solar source region of the disappearance event of 
11 May 1999 and showed that the flows responsible for the event originated 
around 05 May 1999 from a small CH lying adjacent to AR8525.  We examined 
the AR-CH complex at the source region of this 
event using EIT, SXT, and MDI observations.  The observations have 
clearly shown the rapid evolution and changes taking place in the CH lying 
adjacent to AR8525.  The changes are seen to take place in a $\sim$24 hour interval 
starting from 05 May 1999, the approximate launch time of the disappearance 
event.

%
Based on the combined observations, it appears that the rapid evolution seen in the 
CH is due to  a process of interchange reconnection taking place between the
open CH fields and the closed fields from either the bipolar regions to their west 
or other small closed field regions as described above.   The exact magnetic 
topology of the AR-CH region is however, complicated and would require a much more 
detailed study to isolate and pinpoint the reconnection sites and locations of the opposing 
polarities involved. What is clear however, is that the interchange reconnection process 
causes the formation of new bright loops within the CH that can be perceived as a 
progressive constriction taking place across the CH.  Since there is a a high degree 
of correlation between solar wind speed and size of the CH from which it emanates 
\citep{NoK76, Neu94, Wan94, NeF98, KoF99}, we believe that, in this event, the formation 
of the new EUV loops would cause a reduction in the CH area and lead to a suppression of 
CH outflow.  This would then give rise to slower velocity flows form regions that earlier 
produced faster flows, as has been observed in this event.  

As stated above, the rapid changes taking place can be seen to be producing a progressive 
reduction in the area of the CH by causing a clear constriction or pinch-off across it.  The 
observations thus provide support for the mechanism suggested by \cite{JaF05} for causing the 
long lasting low density anomaly or ``{\it{disappearance event}}" at 1 AU.  Since this 
disappearance event was known to have had significant space weather effects \citep{Ros00, 
PaC00}, these observations clearly link the observed effects at 1 AU to a sequence 
of discernible changes taking place in an AR-CH complex on the Sun.  Not considering 
CIRs, these observations thus provide, to the best of our knowledge, the first 
evidence for the so-called Sun-Earth ``transmission-line" arising from non-explosive 
solar events.

Whether or not AR open fields connect to the interplanetary medium to produce solar wind outflows 
has been debated for some years now \citep{KoF99, LuL02, ArH03, SDe03}.  However, the first actual
observations of solar wind outflows from AR open fields located at central meridian and 
lying at the boundary of an AR and a CH have recently been reported \citep{SaK07}.  These 
authors have shown that the observed solar wind outflows came from regions that showed large flux 
expansion factors and low-velocity solar wind, as identified by tomographic IPS observations.  It 
must be noted here that the work by \cite{JaF05} has reported both large flux expansion factors 
and low velocities from the source region of the 11 May 1999 disappearance event.

As opposed to the well-known drivers of space weather phenomena like CME's 
or large flares, disappearance events are not associated with explosive solar 
phenomena.  However, they do produce other observable effects that 
are not fully understood.  For example, \cite{BaJ03} have reported very unusual 
IPS power spectra attributed to high-energy Strahl electrons.  The study of such 
events is therefore important in establishing and understanding solar 
terrestrial relationships in absence of explosive solar events.  With the 
exception of CIR's our observations, as stated earlier, provide the first 
evidence for solar terrestrial connection caused by a non-explosive solar event.  

Solar wind disappearance events constitute extreme deviations from the
average conditions expected in the solar wind at 1 AU.  It would therefore 
be important to continue such studies using both ground and space-based data to 
gain a better understanding of the dynamics and evolution of AR-CH boundary fields.

\begin{acknowledgements}

The authors would like to acknowledge the EIT and MDI consortia for providing 
data in the public domain via the world wide web. SoHO is a project of international 
collaboration between ESA and NASA.  One of the authors, JP, would like to thank 
DAMPT, Cambridge, and STFC for support to initiate this work while DT and HEM acknowledge 
support from STFC.  We thank G. Del Zanna for his comments on the manuscript.  We also
thank the referee for his critical comments and suggestions.
\end{acknowledgements}
\bibliographystyle{aa}

\begin{thebibliography}{33}

\expandafter\ifx\csname natexlab\endcsname\relax\def\natexlab#1{#1}\fi

\bibitem[{{Arge} {et~al.}(2003){Arge}, {Harvey}, {Hudson}, \& {Kahler}}]{ArH03}
{Arge}, C.~N., {Harvey}, K.~L., {Hudson}, H.~S., \& {Kahler}, S.~W. 2003, in
  American Institute of Physics Conference Series, Vol. 679, Solar Wind Ten,
  ed. M.~{Velli}, R.~{Bruno}, F.~{Malara}, \& B.~{Bucci}, 202--205

\bibitem[{{Attrill} {et~al.}(2007){Attrill}, {Harra}, {van Driel-Gesztelyi}, \&
  {D{\'e}moulin}}]{AtH07}
{Attrill}, G.~D.~R., {Harra}, L.~K., {van Driel-Gesztelyi}, L., \&
  {D{\'e}moulin}, P. 2007, \apjl, 656, L101

\bibitem[{{Baker} {et~al.}(2007){Baker}, {van Driel-Gesztelyi}, \&
  {Attrill}}]{BVA07}
{Baker}, D., {van Driel-Gesztelyi}, L., \& {Attrill}, G.~D.~R. 2007,
  Astronomische Nachrichten, 328, 773

\bibitem[{{Balasubramanian} {et~al.}(2003){Balasubramanian}, {Janardhan},
  {Srinivasan}, \& {Ananthakrishnan}}]{BaJ03}
{Balasubramanian}, V., {Janardhan}, P., {Srinivasan}, S., \& {Ananthakrishnan},
  S. 2003, Journal of Geophysical Research (Space Physics), 108, 1121

\bibitem[{{Crooker} {et~al.}(2000){Crooker}, {Shodhan}, {Gosling}, {Simmerer},
  {Lepping}, {Steinberg}, \& {Kahler}}]{CrS00}
{Crooker}, N.~U., {Shodhan}, S., {Gosling}, J.~T., {et~al.} 2000, \grl, 27,
  3769

\bibitem[{{Del Zanna} {et~al.}(2003){Del Zanna}, {Bromage}, \& {Mason}}]{DBM03}
{Del Zanna}, G., {Bromage}, B.~J.~I., \& {Mason}, H.~E. 2003, \aap, 398, 743

\bibitem[{{Delaboudini{\`e}re} {et~al.}(1995){Delaboudini{\`e}re}, {Artzner},
  {Brunaud}, {Gabriel}, {Hochedez}, {Millier}, {Song}, {Au}, {Dere}, {Howard},
  {Kreplin}, {Michels}, {Moses}, {Defise}, {Jamar}, {Rochus}, {Chauvineau},
  {Marioge}, {Catura}, {Lemen}, {Shing}, {Stern}, {Gurman}, {Neupert},
  {Maucherat}, {Clette}, {Cugnon}, \& {van Dessel}}]{DeA95}
{Delaboudini{\`e}re}, J.-P., {Artzner}, G.~E., {Brunaud}, J., {et~al.} 1995,
  \solphys, 162, 291

\bibitem[{{Domingo} {et~al.}(1995){Domingo}, {Fleck}, \& {Poland}}]{DFP95}
{Domingo}, V., {Fleck}, B., \& {Poland}, A.~I. 1995, \solphys, 162, 1

\bibitem[{{Farrugia} {et~al.}(2000){Farrugia}, {Singer}, {Evans},
  {Berdichevsky}, {Scudder}, {Ogilvie}, {Fitzenreiter}, \& {Russell}}]{FaS00}
{Farrugia}, C.~J., {Singer}, H.~J., {Evans}, D., {et~al.} 2000, \grl, 27, 3773

\bibitem[{{Hakamada} \& {Kojima}(1999)}]{HKo99}
{Hakamada}, K. \& {Kojima}, M. 1999, \solphys, 187, 115

\bibitem[{{Harvey} \& {Harvey}(1973)}]{HHa73}
{Harvey}, K. \& {Harvey}, J. 1973, \solphys, 28, 61

\bibitem[{{Janardhan} {et~al.}(2005){Janardhan}, {Fujiki}, {Kojima},
  {Tokumaru}, \& {Hakamada}}]{JaF05}
{Janardhan}, P., {Fujiki}, K., {Kojima}, M., {Tokumaru}, M., \& {Hakamada}, K.
  2005, Journal of Geophysical Research (Space Physics), 110, 8101

\bibitem[{{Janardhan} {et~al.}(2008){Janardhan}, {Fujiki}, {Sawant}, {Kojima},
  {Hakamada}, \& {Krishnan}}]{JaF08}
{Janardhan}, P., {Fujiki}, K., {Sawant}, H.~S., {et~al.} 2008, Journal of
  Geophysical Research (Space Physics), 113, 3102

\bibitem[{{Kojima} {et~al.}(1999){Kojima}, {Fujiki}, {Ohmi}, {Tokumaru},
  {Yokobe}, \& {Hakamada}}]{KoF99}
{Kojima}, M., {Fujiki}, K., {Ohmi}, T., {et~al.} 1999, \jgr, 104, 16993

\bibitem[{{Le} {et~al.}(2000){Le}, {Chi}, {Goedecke}, {Russell}, {Szabo},
  {Petrinec}, {Angelopoulos}, {Reeves}, \& {Chun}}]{LeC00}
{Le}, G., {Chi}, P.~J., {Goedecke}, W., {et~al.} 2000, \grl, 27, 2165

\bibitem[{{Liewer} {et~al.}(2004){Liewer}, {Neugebauer}, \&
  {Zurbuchen}}]{LNZ04}
{Liewer}, P.~C., {Neugebauer}, M., \& {Zurbuchen}, T. 2004, \solphys, 223, 209

\bibitem[{{Lockwood}(2001)}]{Loc01}
{Lockwood}, M. 2001, \nat, 409, 677

\bibitem[{{Luhmann} {et~al.}(2002){Luhmann}, {Li}, {Arge}, {Gazis}, \&
  {Ulrich}}]{LuL02}
{Luhmann}, J.~G., {Li}, Y., {Arge}, C.~N., {Gazis}, P.~R., \& {Ulrich}, R.
  2002, Journal of Geophysical Research (Space Physics), 107, 1154

\bibitem[{{Mandrini} {et~al.}(2007){Mandrini}, {Nakwacki}, {Attrill}, {van
  Driel-Gesztelyi}, {D{\'e}moulin}, {Dasso}, \& {Elliott}}]{MaN07}
{Mandrini}, C.~H., {Nakwacki}, M.~S., {Attrill}, G., {et~al.} 2007, \solphys,
  244, 25

\bibitem[{{Neugebauer}(1994)}]{Neu94}
{Neugebauer}, M. 1994, Space Science Reviews, 70, 319

\bibitem[{{Neugebauer} {et~al.}(1998){Neugebauer}, {Forsyth}, {Galvin},
  {Harvey}, {Hoeksema}, {Lazarus}, {Lepping}, {Linker}, {Mikic}, {Steinberg},
  {von Steiger}, {Wang}, \& {Wimmer-Schweingruber}}]{NeF98}
{Neugebauer}, M., {Forsyth}, R.~J., {Galvin}, A.~B., {et~al.} 1998, \jgr, 103,
  14587

\bibitem[{{Nolte} {et~al.}(1976){Nolte}, {Krieger}, {Timothy}, {Gold},
  {Roelof}, {Vaiana}, {Lazarus}, {Sullivan}, \& {McIntosh}}]{NoK76}
{Nolte}, J.~T., {Krieger}, A.~S., {Timothy}, A.~F., {et~al.} 1976, \solphys,
  46, 303

\bibitem[{{Papitashvili} {et~al.}(2000){Papitashvili}, {Clauer},
  {Christiansen}, {Pilipenko}, {Popov}, {Rasmussen}, {Suchdeo}, \&
  {Watermann}}]{PaC00}
{Papitashvili}, V.~O., {Clauer}, C.~R., {Christiansen}, F., {et~al.} 2000,
  \grl, 27, 3785

\bibitem[{{Richardson} {et~al.}(2000){Richardson}, {Berdichevsky}, {Desch}, \&
  {Farrugia}}]{RiB00}
{Richardson}, I.~G., {Berdichevsky}, D., {Desch}, M.~D., \& {Farrugia}, C.~J.
  2000, \grl, 27, 3761

\bibitem[{{Rostoker}(2000)}]{Ros00}
{Rostoker}, G. 2000, \grl, 27, 3789

\bibitem[{{Sakao} {et~al.}(2007){Sakao}, {Kano}, {Narukage}, {Kotoku}, {Bando},
  {DeLuca}, {Lundquist}, {Tsuneta}, {Harra}, {Katsukawa}, {Kubo}, {Hara},
  {Matsuzaki}, {Shimojo}, {Bookbinder}, {Golub}, {Korreck}, {Su}, {Shibasaki},
  {Shimizu}, \& {Nakatani}}]{SaK07}
{Sakao}, T., {Kano}, R., {Narukage}, N., {et~al.} 2007, Science, 318, 1585

\bibitem[{{Scherrer} {et~al.}(1995){Scherrer}, {Bogart}, {Bush}, {Hoeksema},
  {Kosovichev}, {Schou}, {Rosenberg}, {Springer}, {Tarbell}, {Title},
  {Wolfson}, {Zayer}, \& {MDI Engineering Team}}]{ScB95}
{Scherrer}, P.~H., {Bogart}, R.~S., {Bush}, R.~I., {et~al.} 1995, \solphys,
  162, 129

\bibitem[{{Schrijver} \& {DeRosa}(2003)}]{SDe03}
{Schrijver}, C.~J. \& {DeRosa}, M.~L. 2003, \solphys, 212, 165

\bibitem[{{Stone} {et~al.}(1998){Stone}, {Frandsen}, {Mewaldt}, {Christian},
  {Margolies}, {Ormes}, \& {Snow}}]{StF98}
{Stone}, E.~C., {Frandsen}, A.~M., {Mewaldt}, R.~A., {et~al.} 1998, Space
  Science Reviews, 86, 1

\bibitem[{{Tripathi} {et~al.}(2006){Tripathi}, {Del Zanna}, {Mason}, \&
  {Chifor}}]{TrD06}
{Tripathi}, D., {Del Zanna}, G., {Mason}, H.~E., \& {Chifor}, C. 2006, \aap,
  460, L53

\bibitem[{{Tsuneta} {et~al.}(1991){Tsuneta}, {Acton}, {Bruner}, {Lemen},
  {Brown}, {Caravalho}, {Catura}, {Freeland}, {Jurcevich}, \& {Owens}}]{TsA91}
{Tsuneta}, S., {Acton}, L., {Bruner}, M., {et~al.} 1991, \solphys, 136, 37

\bibitem[{{Usmanov} {et~al.}(2000){Usmanov}, {Goldstein}, \& {Farrell}}]{UGF00}
{Usmanov}, A.~V., {Goldstein}, M.~L., \& {Farrell}, W.~M. 2000, \grl, 27, 3765

\bibitem[{{Wang}(1994)}]{Wan94}
{Wang}, Y.-M. 1994, \apjl, 437, L67

\end{thebibliography}

\end{document}